# Susceptibility to Misinformation about COVID-19 Vaccines: A Signal Detection Analysis

Lea S. Nahon[a,b], Nyx L. Ng[a], and Bertram Gawronski[a]
[a] University of Texas at Austin
[b] University of Marburg

An analysis drawing on Signal Detection Theory suggests that people may fall for misinformation because they are unable to discern true from false information (*truth insensitivity*) or because they tend to accept information with a particular slant regardless of whether it is true or false (*belief bias*). Three preregistered experiments with participants from the United States and the United Kingdom ($N = 961$) revealed that (*i*) truth insensitivity in responses to (mis)information about COVID-19 vaccines differed as a function of prior attitudes toward COVID-19 vaccines; (*ii*) participants exhibited a strong belief bias favoring attitude-congruent information; (*iii*) truth insensitivity and belief bias jointly predicted acceptance of false information about COVID-19 vaccines, but belief bias was a much stronger predictor; (*iv*) cognitive elaboration increased truth sensitivity without reducing belief bias; and (*v*) higher levels of confidence in one's beliefs were associated with greater belief bias. The findings provide insights into why people fall for misinformation, which is essential for individual-level interventions to reduce susceptibility to misinformation.

In May 2022, the Commissioner of the U.S. Food and Drug Administration stated that misinformation has become the leading cause of death in the United States (CNN, 2022). Although vaccines are among the safest and most economical measures in our medical repertoire (Bloom et al., 2014), misinformation about COVID-19 vaccines has contributed to this state of affairs by undermining vaccine acceptance (Enders et al., 2020; Loomba et al., 2021; Schmid & Betsch, 2022). In the current work, we drew on Signal Detection Theory (SDT; Green & Swets, 1966) to better understand why people fall for misinformation about COVID-19 vaccines.

**A Signal Detection Analysis of Misinformation Susceptibility**

From the perspective of SDT, understanding susceptibility to misinformation requires considering the four cases of a 2 × 2 matrix involving judgments of true and false information as either true or false, respectively (Batailler et al., 2022). Using SDT terminology, a judgment of true information as true can be described as a *hit*; a judgment of false information as false can be described as a *correct rejection*; a judgment of true information as false can be described as a *miss*; and a judgment of false information as true can be described as a *false alarm* (see Table 1). The question of why people fall for misinformation is essentially concerned with false alarms: why do people accept false information as true?

According to SDT, one potential reason why people may accept false information as true is that they are unable to distinguish between true and false information (Batailler et al., 2022). In this case, people would show not only a high rate of false alarms, but also a high rate of misses (see Table 1). Statistically, this case would be reflected in low scores on SDT's $d'$ index, which reflects the distance between the distributions of judgments about true and false information along the judgment dimension of veracity. SDT's $d'$ index can be calculated with the following equation:

$$d' = z(H) - z(FA)$$

In this equation, H represents the proportion of hits (i.e., proportion of true information judged true) and FA represents the proportion of false alarms (i.e., proportion of false information judged true), with both H and FA being transformed to a quantile function for a $z$ distribution in a manner such that a proportion of 0.5 is converted to a $z$-score of 0 (Stanislaw & Todorov, 1999).[1] Thus, a $d'$ score of 0 reflects chance-level performance in the identification of true and false information; higher scores reflect greater accuracy in discerning true and false information.

Another reason why people may accept false information as true is that they have a general tendency to judge information as true regardless of whether it is true or false (Batailler et al., 2022). In this case, people would show a high rate of hits in addition to a high rate of false alarms (see Table 1). Statistically, this case would be reflected in low scores on SDT's $c$ index, which reflects the threshold along the judgmental dimension of perceived veracity at which a person decides to switch their decision. SDT's $c$ index can be calculated with the following equation:

$$c = -0.5 \times [z(H) + z(FA)]$$

---

[1] Note that H and FA refer to proportions rather than absolute numbers. Hence, these proportions are mathematically redundant with the proportion of true information judged false (i.e., misses) and the proportion of false information judged false (i.e., correct rejections), respectively, in that $p$(miss) = 1 – $p$(hit) and $p$(correct rejection) = 1 – $p$(false alarm).

A $c$ score of 0 indicates an equal likelihood of information being judged as true or false; $c$ scores greater than zero reflect a higher likelihood of information being judged as false rather than true; and $c$ scores smaller than zero reflect a higher likelihood of information being judged as true rather than false.

Although it seems possible that misinformation susceptibility is driven by a general tendency to accept all information as true, acceptance thresholds likely depend on the content of the relevant information (Batailler et al., 2022). For example, a large body of research suggests that people tend to accept information as true when it is congruent with their beliefs and reject information as false when it is incongruent with their beliefs (for a review, see Brashier & Marsh, 2020). Statistically, this difference would be reflected in a lower score on SDT's $c$ index for belief-congruent information than belief-incongruent information (Batailler et al., 2022). Together, these considerations suggest that people may fall for misinformation because they are unable to discern true from false information (*truth insensitivity*) or because they tend to accept information with a particular slant regardless of whether it is true or false (*belief bias*).

Although a large body of work has provided valuable insights into psychological factors underlying susceptibility to misinformation (for reviews, see Ecker et al., 2022; van der Linden, 2022), an analysis from the perspective of SDT reveals critical drawbacks of this work. For the current purpose, the most significant example is the use of methodological approaches that focus exclusively on the ability to distinguish between true and false information, which has led to confusion about the role of belief bias in responses to misinformation (for discussions, see Batailler et al., 2022; Gawronski, 2021). For example, in research on political misinformation, some researchers concluded that partisan bias is irrelevant for understanding susceptibility to misinformation, because truth discernment for ideology-congruent information is often greater (rather than smaller) than truth discernment for ideology-incongruent information (Pennycook & Rand, 2021a, 2021b). Yet, from the perspective of SDT, partisan bias has nothing to do with truth discernment, but instead involves a lower acceptance threshold for ideology-congruent information than ideology-incongruent information (Batailler et al., 2022; Gawronski, et al., 2023). Importantly, when conceptualized in this manner, partisan bias in responses to political (mis)information is extremely large in terms of current conventions for the interpretation of effect sizes (see Cohen, 1988) and it accounts for a much larger portion of variance in misinformation susceptibility than (in)ability to distinguish between true and false information (Gawronski et al., 2023). Drawing on these considerations, the main goal of the current work was to go beyond the methodological constraints of prior research by using SDT to disentangle the independent contributions of truth insensitivity and belief bias to misinformation susceptibility in the domain of COVID-19-vaccines.

## The Current Work

To this end, we conducted three preregistered experiments. In each experiment, participants judged the veracity of true and false statements about COVID-19 vaccines. Half of the statements had a pro-COVID-19-vaccine slant, and the other half had an anti-COVID-19-vaccine slant. Experiment 1 compared participants with favorable, unfavorable, and neutral attitudes toward COVID-19 vaccines in terms of their truth sensitivity and belief bias, respectively. In addition, we explored whether and to what extent acceptance of false information about COVID-19 vaccines is accounted for by truth insensitivity, belief bias, or both. Expanding on prior work suggesting that higher levels of cognitive elaboration are associated with lower susceptibility to misinformation (e.g., Bago et al., 2020; Pennycook & Rand, 2019), Experiment 2 investigated effects of time pressure on truth sensitivity and belief bias. Experiment 3 aimed to provide deeper insights into the determinants of belief bias by examining (*i*) whether positive self-feelings are negatively associated with belief-congruency bias (as predicted by motivational accounts), and (*ii*) whether self-confidence is positively associated with belief-congruency bias (as predicted by cognitive accounts). Experiments 2 and 3 also included confirmatory analyses to replicate the exploratory results of Experiment 1 for the prediction of misinformation susceptibility by truth insensitivity and belief bias, as well as the obtained differences between vaccine-attitude groups.

## Open Practices

The Institutional Review Board of the authors' institution approved the studies under protocol STUDY00000822. We report how we determined our sample size, all data exclusions, all manipulations, and all measures. Anonymized behavioral data, codebooks, materials, analysis codes, and preregistrations of manipulations, measures, hypotheses, data collection and exclusions, sample size, and analyses can be found at https://osf.io/utk69/. All preregistered analyses are reported in the main article or the Supplemental Materials. Any deviations from the preregistered analysis plan are noted in the article, and exploratory analyses are specified as such. The data were analyzed using IBM SPSS Statistics for Macintosh version 26.0. Cohen's $d$ values were calculated using the spreadsheet provided by Lakens (2013; termed $d_s$ for $t$-tests for independent groups and $d_z$ for $t$-tests for dependent groups). Sensitivity analyses were conducted using

G*Power 3.1.9.6 for Macintosh (Faul et al., 2007). For all power analyses involving mixed ANOVAs, we assumed a correlation between measures of $r = .30$ and used a nonsphericity correction of $\varepsilon = 1$.

## Experiment 1

Experiment 1 had two goals. The first goal was to investigate associations between attitudes toward COVID-19 vaccines and susceptibility to misinformation about COVID-19 vaccines. Based on the notion that prior attitudes can lead people to readily accept information that is congruent with their attitudes and dismiss information that is incongruent with their attitudes (Batailler et al., 2022; Edwards & Smith, 1996; Gawronski et al., 2023; Van Bavel & Pereira, 2018), we examined truth sensitivity and belief bias among participants with favorable, unfavorable, and neutral attitudes toward COVID-19 vaccines.

The second goal was to explore whether and to what extent acceptance of misinformation about COVID-19 vaccines is accounted for by an inability to discern true from false information (*truth insensitivity*) or a tendency to accept information with a particular slant regardless of whether it is true or false (*belief bias*). To this end, we conducted multiple-regression analyses using truth sensitivity and belief bias as simultaneous predictors of acceptance of false information. To ensure statistical independence of predictors and outcome, we calculated scores reflecting truth sensitivity and belief bias based on one half of the data and scores reflecting acceptance of false information based on the other half (and vice versa for a cross-validation of the obtained results). The main question was whether and to what extent acceptance of misinformation is predicted by truth sensitivity and belief bias, respectively.

**Methods**

*Participants and Design*

Data for Experiment 1 were collected in June 2022. We used the crowdsourcing platform Prolific and its prescreening data to separately recruit 150 participants who had reported feeling positively about COVID-19 vaccines, 150 participants who had reported feeling negatively about COVID-19 vaccines, and 150 participants who had reported not having strong opinions either way. The recruitment was based on responses to Prolific's prescreening question *Please describe your attitudes towards the COVID-19 (Coronavirus) vaccines* with the four response options (1) *For (I feel positively about the vaccines)*, (2) *Against (I feel negatively about the vaccines)*, (3) *Neutral (I don't have strong opinions either way)*, (4) *Prefer not to say*. Additional preregistered filters were used to restrict participation to Prolific workers who (*i*) resided in the United States or the United Kingdom, (*ii*) were 18 years old or older, (*iii*) had a minimum approval rate of 90% on prior assignments on Prolific, (*iv*) had completed at least 20 prior assignments on Prolific, and (*v*) were fluent in English. The experiment took approximately 10-15 minutes. Participants were compensated US-$3 for their time. The experiment utilized a 3 (Participant Attitude: favorable vs. unfavorable vs. neutral) × 2 (Statement Accuracy: true vs. false) × 2 (Statement Slant: pro-COVID-19-vaccine vs. anti-COVID-19-vaccine) mixed design with the first factor varying between subjects and the last two factors varying within subjects.

When determining the desired sample size, we anticipated that approximately 10% of participants would be excluded from analyses based on preregistered exclusion criteria (see below). For the preregistered 3 (Participant Attitude, between-subjects) × 2 (Statement Slant, within-subjects) mixed ANOVAs, a sample of $N = 405$ (90% of the full sample) provides a power of 80% to detect small effects of $f = 0.125$ for the main effect of Participant Attitude, $f = 0.083$ for the main effect of Statement Slant, and $f = 0.092$ for the Participant Attitude × Statement Slant interaction.

As preregistered, we ended data collection after 450 participants had been approved credit on Prolific. The number of cases with complete submissions was 451. We used two preregistered criteria to exclude participants with complete submissions from the analyses. First, we excluded 69 participants who failed an attention check.[2] Second, we excluded 59 participants who reported inconsistent attitudes toward COVID-19 vaccines in Prolific's prescreening survey and the measure of attitudes toward COVID-19 vaccines in our experiments.[3] The remaining sample of 323 participants included 117 participants with favorable, 116 participants with unfavorable, and 90 participants with neutral attitudes toward COVID-19 vaccines.

Of the 323 participants in the final sample, 198 identified as female, 119 as male, 2 preferred not to answer, and 4 chose the response option *other*. The age range was 18 to 74 years ($M_{age} = 35.98$, $SD_{age} = 12.60$). Four of the retained participants indicated being American Indian or Alaska Native, 18 Asian, 25 Black, 16 Hispanic, Latino, or Spanish origin, 7 Middle Eastern or North African, 1 Native Hawaiian or Pacific Islander, 272 White, and 7 chose the response option *other*. Of the retained participants, 13 reported having less than a high school diploma or equivalent, 153 a high school diploma or equivalent, 128 an associate or bachelor's degree, 23 a master's degree, and 6 a doctoral degree. Regarding country of residence, 141 participants reported currently residing in the UK and 182 in the US. For the preregistered 3 (Participant Attitude, between-subjects) × 2 (Statement Slant, within-subjects) mixed ANOVAs, the final sample of $N = 323$ provides 80% power to detect small effects of $f = 0.140$ for the main effect of Participant Attitude, $f = 0.093$ for the main effect of Statement Slant, and $f = 0.103$ for the Participant Attitude × Statement Slant interaction.

---

[2] To explore potential effects of selective attrition (Zhou & Fishbach, 2016), we also ran all preregistered analyses without excluding participants who failed the attention check. All findings focal to our main research questions replicated in these analyses.

[3] Exploratory analyses examining truth sensitivity and belief bias among the small number of participants who reported inconsistent attitudes are reported in the Supplemental Materials.

*Materials*

Participants judged the veracity of 20 statements for each of the four statement categories: false pro-COVID-19-vaccine statements (e.g., *If you are vaccinated against COVID-19, you are not going to be hospitalized.*), false anti-COVID-19-vaccine statements (e.g., *Teens are more likely to be hospitalized with myocarditis from the COVID-19 vaccines than to be hospitalized with COVID.*), true pro-COVID-19-vaccine statements (e.g., *Child Covid-19 hospitalizations in the United States rose amid Omicron, especially among children too young to be vaccinated.*), and true anti-COVID-19-vaccine statements (e.g., *COVID-19 vaccines become less effective at preventing severe illness over time.*). The 80 statements were selected from a larger set of statements based on or inspired by online content, such as news articles and social media posts. Some of the statements were left as found online, some were adapted to fit the study, and some were newly created inspired by online statements. We thoroughly screened and fact-checked all selected statements. Details on the materials-selection procedure are reported in the Appendix. Detailed information about the selected statements (e.g., source URL and fact check) and a comprehensive list of all statements can be found at https://osf.io/utk69/.

*Procedure and Measures*

**Country of residence.** Participants first indicated their current country of residence. Only participants who chose *The United Kingdom* or *The United States of America* were admitted for participation in the experiments. If participants selected the response option *Other*, they received information about their ineligibility and the study was terminated.

**Political orientation, COVID-19, and education.** Participants then reported how they consider themselves politically in general, in terms of economic issues, and in terms of social issues on scales ranging from 1 (*Very Liberal*) to 7 (*Very Conservative*). Next, participants were asked about their attitudes toward COVID-19 vaccines with the three response options *For (I feel positively about the vaccines)*, *Against (I feel negatively about the vaccines)*, and *Neutral (I don't have strong opinions either way)*. After that, participants indicated their COVID-19 vaccination status, with the options that they were not vaccinated, had gotten a first dose of a 2-dose series, had gotten a primary series, had gotten a primary series and a booster dose, had gotten two booster doses, or preferred not to answer. Participants also reported on their experience with COVID-19, in that they indicated whether they had tested positive and if so, whether they required professional treatment. Participants were given the option not to answer this question. Participants were then asked about their highest level of education with the response options 1 (*Less than high school diploma, GED, GCE A Level, IB, or equivalent*), 2 (*High school diploma, GED, GCE A Level, IB, or equivalent*), 3 (*Associate or Bachelor's degree*), 4 (*Master's degree*), or 5 (*Doctoral degree*).

**Truth judgments.** For the main statement-judgment task, participants read 80 statements about COVID-19 vaccines that differed in terms of their veracity (true vs. false) and slant (pro-COVID-19-vaccine vs. anti-COVID-19-vaccine). The statements were presented in random order for each participant. Participants were asked to judge whether, to the best of their knowledge, the statement was true or false, with the binary response options *True* and *False*. Each statement was presented on a separate page. Participants had unlimited time to provide an answer. Responses were required for all statements.

**Demographics and attention check.** In the final part of the experiment, participants completed demographic questions about their gender, age, and racial/ethnic identity. Next, participants completed an attention check, in which they were asked not to select any of the response options. Participants who selected one or more of the response options were classified as having failed the attention check. After being debriefed, participants were redirected to Prolific for payment. The complete survey with the verbatim wording of all questions and response options can be found at https://osf.io/utk69/.

*Data Aggregation*

Following our preregistered data-aggregation plan, we computed hit rates as the proportion of true statements judged as true and false-alarm rates as the proportion of false statements judged as true. In cases where the proportion of *true* judgments was either 0 or 1, we followed recommendations by MacMillan and Creelman (2004) and converted values of 0 to $1/(2N)$ and values of 1 to $1-1/(2N)$, where $N$ is the number of trials per statement category (i.e., $N = 20$). Both hit and false-alarm rates were calculated separately for statements with a pro-COVID-19-vaccine slant and statements with an anti-COVID-19-vaccine slant. Hit and false-alarm rates were then used to calculate SDT scores reflecting truth sensitivity ($d'$) and acceptance threshold ($c$) for statements with a pro-COVID-19-vaccine slant and for statements with an anti-COVID-19-vaccine slant, respectively. Overall truth-sensitivity and overall acceptance-threshold scores were calculated by computing the mean $d'$ and $c$ scores across pro- and anti-COVID-19-vaccine statements. Anti-COVID-19-vaccine belief bias was computed as the difference between acceptance thresholds for pro- versus anti-COVID-19-vaccine statements, with higher scores reflecting a lower threshold for accepting statements with an anti-COVID-19-vaccine slant compared to statements with a pro-COVID-19-vaccine slant.

Results

Non-preregistered descriptive analyses revealed that participants showed a high ability to differentiate between true and false information about COVID-19 vaccines, in that the average *d'* score at the sample level was positive and significantly different from zero ($M$ = 1.47, $SD$ = 0.67), $t(322)$ = 39.58, $p$ < .001, $d$ = 2.20. On average, participants were more likely to reject than to accept information about COVID-19 vaccines, which is reflected in a positive overall *c* score that significantly differed from zero ($M$ = 0.21, $SD$ = 0.24), $t(322)$ = 15.57, $p$ < .001, $d$ = 0.87. At the sample level, participants showed an anti-COVID-19-vaccine belief bias, in that *c* scores were significantly lower for anti-COVID-19-vaccine than pro-COVID-19-vaccine information ($M$s = -0.02 vs. 0.43, respectively), $F(1, 320)$ = 83.43, $p$ < .001, $\eta^2_p$ = .21. Truth sensitivity and anti-COVID-19-vaccine belief bias were negatively correlated, in that greater truth sensitivity was associated with a lower anti-COVID-19-vaccine belief bias ($r$ = -.55, $p$ < .001).[4]

*Truth Sensitivity*

Following our preregistered analysis plan, *d'* scores were submitted to a 3 (Participant Attitude) × 2 (Statement Slant) mixed ANOVA with the first variable as a between-subjects factor and the second variable as a within-subjects factor. A significant main effect of Participant Attitude revealed that participants with different attitudes toward COVID-19 vaccines differed in terms of their truth sensitivity, $F(2, 320)$ = 52.08, $p$ < .001, $\eta^2_p$ = .25 (see Figure 1, top-left panel). Preregistered follow-up two-tailed *t*-tests for independent groups revealed that participants with favorable attitudes toward COVID-19 vaccines showed a higher ability to discern true from false information about COVID-19 vaccines compared to participants with neutral attitudes, $t(205)$ = 5.04, $p$ < .001, $d$ = 0.71, and participants with unfavorable attitudes, $t(231)$ = 10.49, $p$ < .001, $d$ = 1.37. Moreover, participants with neutral attitudes showed greater truth sensitivity than participants with unfavorable attitudes, $t(204)$ = -4.46, $p$ < .001, $d$ = 0.63. Non-preregistered exploratory analyses revealed that the main effect of Participant Attitude remained statistically significant after controlling for participants' country of residence, gender, racial/ethnic identity, political orientation, age, and education, $F(2, 306)$ = 27.03, $p$ < .001.

A significant interaction between Participant Attitude and Statement Slant further indicated that participants with different attitudes toward COVID-19-vaccines differed in terms of their truth sensitivity for pro- and anti-COVID-19-vaccine information, $F(2, 320)$ = 8.17, $p$ < .001, $\eta^2_p$ = .05 (see Figure 1, top-left panel). Participants with favorable attitudes showed greater truth sensitivity for pro- compared to anti-COVID-19-vaccine information, $t(116)$ = 3.37, $p$ = .001, $d_z$ = 0.31. Conversely, participants with unfavorable attitudes showed greater truth sensitivity for anti- compared to pro-COVID-19-vaccine information, $t(115)$ = -2.58, $p$ = .011, $d_z$ = 0.24. Truth sensitivity for pro- and anti-COVID-19-vaccine information did not significantly differ among participants with neutral attitudes, $t(89)$ = 0.33, $p$ = .741, $d_z$ = 0.03.

Nevertheless, the pattern of differences between the three attitude groups was consistent across the two types of information. For pro-COVID-19-vaccine information, truth sensitivity was greater among participants with favorable attitudes compared to participants with unfavorable attitudes, $t(231)$ = 11.70, $p$ < .001, $d$ = 1.53, and participants with neutral attitudes, $t(205)$ = 5.18, $p$ < .001, $d$ = 0.73. Truth sensitivity among participants with neutral attitudes was greater compared to participants with unfavorable attitudes, $t(204)$ = -5.21, $p$ < .001, $d$ = 0.73. For anti-COVID-19-vaccine information, truth sensitivity was again greater among participants with favorable attitudes compared to participants with unfavorable attitudes, $t(231)$ = 6.58, $p$ < .001, $d$ = 0.86, and participants with neutral attitudes, $t(205)$ = 3.16, $p$ = .002, $d$ = 0.44. Truth sensitivity among participants with neutral attitudes was again greater compared to participants with unfavorable attitudes, $t(204)$ = -2.79, $p$ = .006, $d$ = 0.39. Thus, regardless of information slant (i.e., pro-COVID-19-vaccine vs. anti-COVID-19-vaccine), participants with favorable attitudes toward COVID-19 vaccines showed greater truth sensitivity than participants with neutral attitudes, who in turn showed greater truth sensitivity than participants with unfavorable attitudes.[5]

*Acceptance Threshold*

Following our preregistered analysis plan, *c* scores were submitted to a 3 (Participant Attitude) × 2 (Statement Slant) mixed ANOVA with the first variable as a between-subjects factor and the second variable as a within-subjects factor. A significant main effect of Participant Attitude revealed that participants with different attitudes toward COVID-19 vaccines differed in their overall acceptance-thresholds, $F(2, 320) = 7.72$, $p = .001$, $\eta^2_p = .05$ (see Figure 1, top-right panel). Participants with unfavorable attitudes had a significantly higher acceptance threshold than participants with favorable attitudes, $t(231) = -4.21$, $p < .001$, $d = 0.55$, and participants with neutral attitudes had a significantly higher acceptance threshold than participants with favorable attitudes, $t(205) = -2.64$, $p = .009$, $d = 0.37$. Participants with neutral attitudes did not significantly differ from participants with unfavorable attitudes in terms of their acceptance thresholds, $t(204) = 0.74$, $p = .461$, $d = 0.10$.

---

[4] Non-preregistered exploratory analyses on demographic differences are reported in the Supplemental Materials.
[5] Exploratory analyses that additionally included country (US vs. UK) as a factor in the preregistered ANOVA for truth sensitivity revealed no reliable differences between participants from the US and the UK (i.e., main effect of country or interactions with country) that replicated across the three experiments.

A significant interaction between Participant Attitude and Statement Slant further indicated that participants with different attitudes toward COVID-19 vaccines differed in their acceptance thresholds for pro- versus anti-COVID-19-vaccine information, $F(2, 320) = 181.74$, $p < .001$, $\eta^2_p = .53$ (see Figure 1, top-right panel). Participants with favorable attitudes had a significantly lower acceptance threshold for pro-COVID-19-vaccine information than anti-COVID-19-vaccine information, and thus showed a pro-COVID-19-vaccine belief bias, $t(116) = -10.32$, $p < .001$, $d_z = 0.95$. In contrast, participants with unfavorable attitudes had a significantly lower acceptance threshold for anti-COVID-19-vaccine information than pro-COVID-19-vaccine information, and thus showed an anti-COVID-19-vaccine belief bias, $t(115) = 15.37$, $p < .001$, $d_z = 1.43$. Participants with neutral attitudes also showed an anti-COVID-19-vaccine belief bias, in that they had a significantly lower acceptance threshold for anti-COVID-19-vaccine information than pro-COVID-19-vaccine information, $t(89) = 3.70$, $p < .001$, $d_z = 0.39$.[6] Non-preregistered exploratory analyses revealed that the interaction between Participant Attitude and Statement Slant remained statistically significant after controlling for country of residence, gender, racial/ethnic identity, political orientation, age, and education, $F(2, 320) = 181.74$, $p < .001$.[7]

*Acceptance of False Information*

To investigate the extent to which acceptance of false information about COVID-19 vaccines is predicted by truth insensitivity and belief bias, we conducted preregistered exploratory analyses regressing acceptance of false information (i.e., false-alarm rate) onto overall truth-sensitivity and anti-COVID-19-vaccine belief bias as simultaneous predictors (for more details, see Supplemental Materials). Because false-alarm rates are used to compute overall truth-sensitivity and anti-COVID-19-vaccine belief bias scores, we ensured mathematical independence by using different subsets of data for the calculation of predictor and outcome scores: one subset included responses to odd-numbered items in our data sets and the other subset included responses to even-numbered items (see Gawronski et al., 2023). We then conducted two conceptually equivalent regression analyses. First, false-alarm rate on odd-numbered items was regressed onto overall truth-sensitivity and anti-COVID-19-vaccine belief bias computed based on even-numbered items. Second, false-alarm rate on even-numbered items was regressed onto overall truth-sensitivity and anti-COVID-19-vaccine belief bias computed based on odd-numbered items. Regression analyses were conducted separately for the acceptance of anti-COVID-19-vaccine misinformation and the acceptance of pro-COVID-19-vaccine misinformation.

The results of the multiple-regression analyses are presented in Table 2. Truth sensitivity showed a reliable negative association with acceptance of false information about COVID-19 vaccines regardless of whether the false information had a pro- or anti-COVID-19-vaccine slant. Anti-COVID-19-vaccine belief bias also predicted acceptance of false information, with standardized regression coefficients that were approximately twice as large as those obtained for truth sensitivity. Anti-COVID-19-vaccine belief bias showed a reliable positive association with acceptance of anti-COVID-19-vaccine misinformation and a reliable negative association with acceptance of pro-COVID-19-vaccine misinformation. The latter finding reflects the fact that anti-COVID-19-vaccine belief bias involves a rejection of all positive information about COVID-19 vaccines, including positive misinformation. All effects replicated regardless of whether the outcome variable was calculated based on odd-numbered items and the predictors based on even-numbered items, or vice versa.[8]

**Discussion**

Experiment 1 obtained three sets of important findings. First, truth sensitivity differed as a function of prior attitudes, in that participants with favorable attitudes toward COVID-19 vaccines showed the highest ability in discerning true and false information about COVID-19 vaccines, while participants with unfavorable attitudes toward COVID-19 vaccines showed the lowest ability. Participants with neutral attitudes showed truth-sensitivity levels in-between the two groups. Second, while participants with favorable attitudes showed a lower acceptance threshold for pro-COVID-19-vaccine information than anti-COVID-19-vaccine information (i.e., pro-COVID-19-vaccine belief bias), participants with neutral and unfavorable attitudes showed the reverse pattern (i.e., anti-COVID-19-vaccine belief bias). Third, although acceptance of false information about COVID-19 vaccines was predicted by both truth sensitivity and belief bias, the obtained associations with belief bias were substantially larger compared to the associations with truth sensitivity. Together, these results suggest that, while truth insensitivity is an important determinant of susceptibility to misinformation about COVID-19 vaccines, belief bias plays an equally, if not more important role than truth insensitivity. Expanding on this conclusion, Experiments 2 and 3 aimed to provide deeper insights into the determinants of truth insensitivity and belief bias.

---

[6] Follow-up tests comparing acceptance thresholds for each pair of the three Participant Attitude groups for pro-vaccine versus anti-vaccine information, respectively, are reported in the Supplemental Materials.

[7] Exploratory analyses that additionally included country (US vs. UK) as a factor in the preregistered ANOVA for acceptance threshold revealed no reliable differences between participants from the US and the UK (i.e., main effect of country or interactions with country) that replicated across the three experiments.

[8] Because truth sensitivity and anti-COVID-19-vaccine belief bias were negatively correlated, we computed variance inflation factors (VIFs) to rule out potential problems with multicollinearity. The VIFs suggest that multicollinearity is not problematic in the current data; predictors even-numbered items: VIF = 1.38, predictors odd-numbered items: VIF = 1.54.

## Experiment 2

Experiment 2 had two goals. The first goal was to replicate the main findings of Experiment 1 on the obtained differences between vaccine-attitude groups and the prediction of misinformation susceptibility by truth insensitivity and belief bias. The second goal was to investigate effects of cognitive elaboration on responses to (mis)information about COVID-19 vaccines. Prior research suggests that greater cognitive elaboration is associated with greater discernment between true and false information (e.g., Bago et al., 2020; Pennycook & Rand, 2019). In Experiment 2, we expanded on these findings by testing whether fast versus slow processing of (mis)information about COVID-19 vaccines influences truth sensitivity and belief bias, respectively (see Kahneman, 2011; Pennycook, 2023). To that end, we restricted response times for half of the participants to 7 seconds per statement, whereas the other half had unlimited time to respond.

### Methods

*Participants and Design*

Data for Experiment 2 were collected in August 2022. We used the crowdsourcing platform Prolific and its prescreening data to separately recruit 200 participants who had reported feeling positively about COVID-19 vaccines, 200 participants who had reported feeling negatively about COVID-19 vaccines, and 200 participants who had reported not having strong opinions either way. The recruitment was based on responses to Prolific's prescreening question *Please describe your attitudes towards the COVID-19 (Coronavirus) vaccines* with the four response options (1) *For (I feel positively about the vaccines)*, (2) *Against (I feel negatively about the vaccines)*, (3) *Neutral (I don't have strong opinions either way)*, (4) *Prefer not to say*. In addition to using the same recruitment filters as in Experiment 1, we restricted participation to Prolific workers who had not participated in Experiment 1. The experiment took approximately 10-15 minutes. Participants were compensated US-$3 for their time. The experiment utilized a 3 (Participant Attitude: favorable vs. unfavorable vs. neutral) × 2 (Cognitive Elaboration: low vs. high) × 2 (Statement Accuracy: true vs. false) × 2 (Statement Slant: pro-COVID-19-vaccine vs. anti-COVID-19-vaccine) mixed design with the first two factors varying between subjects and the last two factors varying within subjects.

When determining the desired sample size, we anticipated that approximately 10% of participants would be excluded from analyses based on preregistered exclusion criteria (see below). For the preregistered 3 (Participant Attitude, between-subjects) × 2 (Cognitive Elaboration, between-subjects) × 2 (Statement Slant, within-subjects) mixed ANOVAs, a sample of $N = 540$ (90% of the sample) provides a power of 80% to detect small effects of $f = 0.125$ for a between-subjects main effect, $f = 0.071$ for a within-subjects main effect, and $f = 0.092$ for a within-between interaction. For the preregistered linear multiple-regression analyses with two predictors, a sample of $N = 540$ provides a power of 80% to detect a small effect of $f^2 = 0.015$ of one predictor.

As preregistered, we ended data collection after 600 participants had been approved credit on Prolific. The number of cases with complete submissions was 601. We used three preregistered criteria to exclude participants with complete submissions from the analyses. First, we excluded 160 participants who failed our attention check. Second, we excluded 14 participants in the low-elaboration condition because they did not respond within 7 seconds for more than five statements within one or more of the four statement categories.[9] Third, we excluded 79 participants who reported inconsistent attitudes toward COVID-19 vaccines in Prolific's prescreening survey and the measure of attitudes toward COVID-19 vaccines in our experiments.[10] The remaining sample of 348 participants included 133 participants with favorable attitudes ($n = 61$ in the low-elaboration condition; $n = 72$ in the high-elaboration condition), 126 participants with unfavorable attitudes (*n* = 58 in the low-elaboration condition; *n* = 68 in the high-elaboration condition), and 89 participants with neutral attitudes (*n* = 36 in the low-elaboration condition; *n* = 53 in the high-elaboration condition).

---

[9] To explore potential effects of selective attrition (Zhou & Fishbach, 2016), we also ran all preregistered analyses without excluding participants who did not pass the attention check or failed to respond within 7 seconds for more than five statements within one or more of the four statement categories. All findings focal to our main research questions replicated in these analyses. One non-focal difference occurred, in that participants with neutral attitudes showed significantly greater truth sensitivity for pro- compared to anti-COVID-19-vaccine information.

[10] Exploratory analyses examining truth sensitivity and belief bias among the small number of participants who reported inconsistent attitudes are reported in the Supplemental Materials.

Of the 348 participants in the final sample, 236 identified as female, 110 as male, and 2 chose the response option *other*. The age range was 18 to 80 years ($M_{age}$ = 39.22, $SD_{age}$ = 12.67). One of the retained participants indicated being American Indian or Alaska Native, 21 Asian, 14 Black, 8 Hispanic, Latino, or Spanish origin, 2 Middle Eastern or North African, 304 White, and 7 chose the response option *other*. Of the retained participants, 13 reported having less than a high school diploma or equivalent, 153 a high school diploma or equivalent, 138 an associate or bachelor's degree, 37 a master's degree, and 7 a doctoral degree. Regarding country of residence, 299 participants reported currently residing in the UK and 49 in the US. For the preregistered 3 (Participant Attitude, between-subjects) × 2 (Cognitive Elaboration, between-subjects) × 2 (Statement Slant, within-subjects) mixed ANOVAs, the final sample of *N* = 348 provides 80% power to detect small effects of *f* = 0.156 for a between-subjects main effect, *f* = 0.089 for a within-subjects main effect, and *f* = 0.115 for a within-between interaction. For the preregistered multiple-regressions with two predictors, the final sample of *N* = 348 provides 80% power to detect a small effect of $f^2$ = 0.023 of one predictor.

*Materials, Procedure, and Measures*

The materials, procedure, and measures were identical to Experiment 1, the only difference being that Experiment 2 included a manipulation of cognitive elaboration. Participants in the low-elaboration condition were asked to respond to the questions based on their initial reactions to each statement. They were further informed that they had a 7-second time limit to provide a response and should therefore provide their answer to the question as quickly as possible. To reiterate these instructions, the note *You have 7 seconds to read and respond to the following statement* was presented above each statement. Each statement and its accompanying question were presented to participants for 7 seconds; the experiment automatically advanced to the next screen after the 7 second time limit. Participants in the high-elaboration condition were asked to respond to the questions based on careful consideration to each statement. They were further informed that they had unlimited time to provide a response and should therefore provide their answer to the question only after thinking carefully. To reiterate these instructions, the note *Please read the statement carefully and take a moment to think about your answer* was presented above each statement. Unlike participants in the low-elaboration condition, participants in the high-elaboration condition had unlimited time to provide an answer. Responses were required for all statements except for the truth judgments in the low-elaboration condition (because this condition included a time limit to respond).

**Results**

As preregistered, data aggregation followed the procedures outlined in Experiment 1. Non-preregistered descriptive analyses revealed that participants showed a high ability to differentiate between true and false information about COVID-19 vaccines, in that the average *d'* score in the total sample was positive and significantly different from zero (*M* = 1.37, *SD* = 0.69), *t*(347) = 36.93, *p* < .001, *d* = 1.98. Further analyses revealed that, on average, participants were more likely to reject than to accept information about COVID-19 vaccines, which is reflected in a positive *c* score that significantly differed from zero (*M* = 0.21, *SD* = 0.25), *t*(347) = 16.16, *p* < .001, *d* = 0.87. At the sample level, participants showed an anti-COVID-19-vaccine belief bias, in that *c* scores were significantly lower for anti-COVID-19-vaccine than pro-COVID-19-vaccine information (*M*s = -0.10 vs. 0.53, respectively), *F*(1, 342) = 227.73, *p* < .001, $\eta^2_p$ = .40. Truth sensitivity and anti-COVID-19-vaccine belief bias were negatively correlated, in that greater truth sensitivity was associated with a lower anti-COVID-19-vaccine belief bias (*r* = -.56, *p* < .001). Non-preregistered exploratory analyses revealed that the cognitive-elaboration manipulation was successful, in that participants in the high-elaboration condition took significantly longer to complete the study than participants in the low-elaboration condition (*M*s = 894.95 vs. 645.39 seconds, respectively), *t*(346) = -5.99, *p* < .001, *d* = 0.65.[11]

*Truth Sensitivity*

Following our preregistered analysis plan, *d'* scores were submitted to a 3 (Participant Attitude) × 2 (Cognitive Elaboration) × 2 (Statement Slant) mixed ANOVA with the first two variables as between-subjects factors and the last variable as a within-subjects factor. A significant main effect of Participant Attitude revealed that participants with different attitudes toward COVID-19 vaccines differed in terms of their truth sensitivity, *F*(2, 342) = 82.77, *p* < .001, $\eta^2_p$ = .33 (see Figure 1, middle-left panel). Replicating the results of Experiment 1, participants with favorable attitudes toward COVID-19 vaccines showed a higher ability in discerning true from false information about COVID-19 vaccines compared to participants with neutral attitudes, *t*(220) = 4.90, *p* < .001, *d* = 0.67, and participants with unfavorable attitudes, *t*(257) = 12.94, *p* < .001, *d* = 1.61.

---

[11] Non-preregistered exploratory analyses on demographic differences are reported in the Supplemental Materials.

Moreover, participants with neutral attitudes showed significantly greater truth sensitivity than participants with unfavorable attitudes, $t(213) = -6.25$, $p < .001$, $d = 0.87$. Non-preregistered exploratory analyses revealed that the main effect of Participant Attitude remained statistically significant after controlling for participants' country of residence, gender, racial/ethnic identity, political orientation, age, and education, $F(2, 329) = 49.15$, $p < .001$.

In addition to the main effect of Participant Attitude, there was a significant main effect of Statement Slant, indicating that participants showed greater truth sensitivity for pro- compared to anti-COVID-19-vaccine information, $F(1, 342) = 10.72$, $p = .001$, $\eta^2_p = .03$ (see Figure 1, middle-left panel). The two main effects were qualified by a significant two-way interaction between Participant Attitude and Statement Slant, indicating that participants with different attitudes toward COVID-19-vaccines differed in terms of their truth sensitivity for pro- and anti-COVID-19-vaccine information, $F(2, 342) = 5.36$, $p = .005$, $\eta^2_p = .03$ (see Figure 1, middle-left panel). Replicating the results of Experiment 1, participants with favorable attitudes showed greater truth sensitivity for pro- compared to anti-COVID-19-vaccine information, $t(132) = 3.59$, $p < .001$, $d_z = 0.31$. Yet, different from the results of Experiment 1, truth sensitivity for the two kinds of information did not significantly differ among participants with unfavorable attitudes, $t(125) = -0.89$, $p = .376$, $d_z = 0.08$. Truth sensitivity for pro- and anti-COVID-19-vaccine information also did not significantly differ among participants with neutral attitudes, $t(88) = 1.90$, $p = .061$, $d_z = 0.20$.

Nevertheless, the pattern of differences between the three attitude groups was again consistent across the two types of statements, replicating the results of Experiment 1. For pro-COVID-19-vaccine information, truth sensitivity was greater among participants with favorable attitudes compared to participants with unfavorable attitudes, $t(257) = 14.37$, $p < .001$, $d = 1.79$, and participants with neutral attitudes, $t(220) = 4.71$, $p < .001$, $d = 0.64$. Truth sensitivity among participants with neutral attitudes was greater compared to participants with unfavorable attitudes, $t(213) = -7.44$, $p < .001$, $d = 1.03$. For anti-COVID-19-vaccine information, truth sensitivity was greater among participants with favorable attitudes compared to participants with unfavorable attitudes, $t(257) = 8.55$, $p < .001$, $d = 1.06$, and participants with neutral attitudes, $t(220) = 3.68$, $p < .001$, $d = 0.50$. Truth sensitivity among participants with neutral attitudes was greater compared to participants with unfavorable attitudes, $t(213) = -3.95$, $p < .001$, $d = 0.55$. Thus, regardless of information slant (i.e., pro-COVID-19-vaccine vs. anti-COVID-19-vaccine), participants with favorable attitudes toward COVID-19 vaccines showed greater truth sensitivity than participants with neutral attitudes, who in turn showed greater truth sensitivity than participants with unfavorable attitudes.

In addition to replicating the differences between attitude groups found in Experiment 1, preregistered confirmatory analyses revealed a significant main effect of Cognitive Elaboration, indicating that truth sensitivity was significantly greater in the high-elaboration condition than in the low-elaboration condition, $F(1, 342) = 14.60$, $p < .001$, $\eta^2_p = .04$.[12] However, this main effect was qualified by an unexpected two-way interaction between Cognitive Elaboration and Statement Slant, $F(1, 342) = 21.88$, $p < .001$, $\eta^2_p = .06$. Non-preregistered post-hoc analyses revealed that truth sensitivity was significantly greater in the high-elaboration condition compared to the low-elaboration condition only for anti-COVID-19-vaccine information, $t(346) = 4.89$, $p < .001$, $d = 0.53$, but not for pro-COVID-19-vaccine information, $t(346) = 0.92$, $p = .361$, $d = 0.10$ (see Figure 2). The two-way interaction between Participant Attitude and Cognitive Elaboration was not statistically significant, $F(2, 342) = 0.19$, $p = .827$, $\eta^2_p = .001$, as was the three-way interaction between Participant Attitude, Cognitive Elaboration, and Statement Slant $F(2, 342) = 0.36$, $p = .696$, $\eta^2_p = .002$. Non-preregistered exploratory analyses revealed that the main effect of Cognitive Elaboration and the two-way interaction between Cognitive Elaboration and Statement Slant remained statistically significant after controlling for participants' country of residence, gender, racial/ethnic identity, political orientation, age, and education, main effect: $F(1, 329) = 18.79$, $p < .001$, two-way interaction: $F(1, 342) = 21.88$, $p < .001$.

*Acceptance Threshold*

Following our preregistered analysis plan, *c* scores were submitted to a 3 (Participant Attitude) × 2 (Cognitive Elaboration) × 2 (Statement Slant) mixed ANOVA with the first two variables as between-subjects factors and the last variable as a within-subjects factor. A significant two-way interaction between Participant Attitude and Statement Slant indicated that participants with different attitudes toward COVID-19 vaccines differed in their acceptance thresholds for pro- versus anti-COVID-19-vaccine information, $F(2, 342) = 294.94$, $p < .001$, $\eta^2_p = .63$ (see Figure 1, middle-right panel). Replicating the results of Experiment 1, participants with favorable attitudes had a significantly lower acceptance threshold for pro-COVID-19-vaccine information than anti-COVID-19-vaccine information, and thus showed a pro-COVID-19-vaccine belief bias, $t(132) = -7.33$, $p < .001$, $d_z = 0.64$. Conversely, participants with unfavorable attitudes had a significantly lower acceptance threshold for anti-COVID-19-vaccine information than pro-COVID-19-vaccine information, and thus showed an anti-COVID-19-vaccine belief bias, $t(125) = 23.45$, $p < .001$, $d_z = 2.09$. Participants with neutral attitudes also showed an anti-COVID-19-vaccine belief bias, in that they had a significantly lower acceptance threshold for anti-COVID-19-vaccine information than pro-COVID-19-vaccine information, $t(88) = 4.20$, $p < .001$, $d_z = 0.45$.[13] Non-preregistered exploratory analyses revealed that the two-way interaction between Participant Attitude and Statement Slant remained statistically significant after controlling for country of residence, gender, racial/ethnic identity, political orientation, age, and education, $F(2, 342) = 294.94$, $p < .001$.

---

[12] Using the data from Experiment 1 as a baseline, non-preregistered exploratory analyses revealed that truth sensitivity was significantly lower in the low-elaboration condition compared to baseline, $t(476) = -3.44$, $p = .001$, $d = 0.34$. Truth sensitivity in the high-elaboration condition did not significantly differ from baseline, $t(514) = 0.16$, $p = .876$, $d = 0.01$.

Cognitive elaboration had no significant effect on belief bias, as reflected in a non-significant two-way interaction between Cognitive Elaboration and Statement Slant, $F(1, 342) = 2.32$, $p = .128$, $\eta^2_p = .01$, and a non-significant three-way interaction between Participant Attitude, Cognitive Elaboration, and Statement Slant, $F(2, 342) = 1.07$, $p = .345$, $\eta^2_p = .01$. Cognitive elaboration neither had a significant main effect of its own, $F(1, 342) = 0.15$, $p = .699$, $\eta^2_p < .001$, nor did it interact with Participant Attitude, $F(2, 342) = 0.37$, $p = .693$, $\eta^2_p = .002$. The two-way interaction of Cognitive Elaboration and Statement Slant was not statistically significant for all three Participant Attitude groups, favorable: $F(1, 131) = 0.14$, $p = .711$, $\eta^2_p = .001$, unfavorable: $F(1, 124) = 1.52$, $p = .220$, $\eta^2_p = .01$, neutral: $F(1, 87) = 1.77$, $p = .186$, $\eta^2_p = .02$. Non-preregistered exploratory analyses revealed that all effects on acceptance threshold involving Cognitive Elaboration remained non-significant after controlling for participants' country of residence, gender, racial/ethnic identity, political orientation, age, and education (all $F$s < 2.33, all $p$s > .12).

*Acceptance of False Information*

Preregistered confirmatory analyses regressing acceptance of false information (i.e., false-alarm rate) onto overall truth-sensitivity and anti-COVID-19-vaccine belief bias as simultaneous predictors followed the procedures outlined in Experiment 1. The results of the multiple-regression analyses are presented in Table 2. Replicating the results of Experiment 1, truth sensitivity showed a reliable negative association with acceptance of false information about COVID-19 vaccines regardless of whether the false information had a pro- or anti-COVID-19-vaccine slant. Also replicating the results of Experiment 1, anti-COVID-19-vaccine belief bias showed a reliable positive association with acceptance of anti-COVID-19-vaccine misinformation and a reliable negative association with acceptance of pro-COVID-19-vaccine misinformation. Standardized regression coefficients for the obtained associations with belief bias were again approximately twice as large as those obtained for truth sensitivity. All effects replicated regardless of whether the outcome variable was calculated based on odd-numbered items and the predictors based on even-numbered items, or vice versa.[14]

**Discussion**

Experiment 2 obtained three sets of important findings. First, preregistered confirmatory analyses replicated the differences between vaccine-attitude groups obtained in Experiment 1, in that (*i*) participants with favorable attitudes toward COVID-19 vaccines showed the highest levels of truth sensitivity and participants with unfavorable attitudes showed the lowest levels, with participants with neutral attitudes showing truth-sensitivity levels in-between the two groups, and (*ii*) participants with favorable attitudes toward COVID-19 vaccines showed a pro-COVID-19-vaccine belief bias, whereas participants with neutral and unfavorable attitudes showed an anti-COVID-19-vaccine belief bias. Second, preregistered confirmatory analyses replicated the predictive relations obtained in Experiment 1, in that acceptance of false information about COVID-19 vaccines was jointly predicted by both truth sensitivity and belief bias, with belief bias being the stronger predictor. Third, preregistered confirmatory analyses revealed that cognitive elaboration increased truth sensitivity without reducing belief bias. Yet, unexpectedly, the effect of cognitive elaboration on truth sensitivity was limited to anti-COVID-19-vaccine information and did not generalize to pro-COVID-19-vaccine information. We will return to this unexpected finding in the General Discussion.

**Experiment 3**

The results of Experiment 2 suggest that elaborate thinking can (at least partially) increase truth sensitivity in responses to (mis)information about COVID-19 vaccines. However, elaborate thinking was ineffective in reducing belief bias. Because the results of Experiments 1 and 2 suggest that belief bias is a much stronger predictor of misinformation susceptibility than truth insensitivity, it seems important to understand the determinants of belief bias. Experiment 3 aimed to address this question by testing predictions of two competing accounts of belief bias.

---

[13] Follow-up tests comparing acceptance thresholds for each pair of the participant groups for pro-vaccine versus anti-vaccine information, respectively, are reported in the Supplemental Materials.

[14] Because truth sensitivity and anti-COVID-19-vaccine belief bias were negatively correlated, we computed variance inflation factors (VIFs) to rule out potential problems with multicollinearity. The VIFs suggest that multicollinearity is not problematic in the current data; predictors even-numbered items: VIF = 1.49, predictors odd-numbered items: VIF = 1.56.

According to motivational accounts, belief bias is a product of motivated reasoning (Ditto et al., in press; Kunda, 1990; Kruglanski et al., 2020). A central assumption underlying these accounts is that people have a deeply rooted need to feel good about themselves (e.g., Alicke & Sedikides, 2009), which gives rise to a desire to support and protect subjectively important beliefs. On the one hand, the need for positive self-regard leads people to readily accept information that supports their personal beliefs, because the self-validation implied by belief-congruent information elicits positive feelings about the self. On the other hand, the need for positive self-regard leads people to readily reject information that questions their personal beliefs, because the self-threat implied by belief-incongruent information elicits negative feelings about the self. Moreover, from a homeostatic view, this tendency should be less pronounced when the need for positive self-regard is satiated, and it should be more pronounced when the need for positive self-regard is deprived (Sherman & Cohen, 2006). Hence, people who feel relatively positive about themselves should be less prone to readily accept belief-congruent information and reject belief-incongruent information because they do not need to regulate their self-feelings by seeking self-validation and avoiding self-threat (Liberman & Chaiken, 1992; Sherman et al., 2000). Conversely, people who feel less positive about themselves should have a stronger tendency to readily accept belief-congruent information and reject belief-incongruent information because doing so helps to elevate positive feelings about the self (Liberman & Chaiken, 1992; Sherman et al., 2000). Together, these assumptions suggest that positive self-feelings should be negatively associated with belief bias favoring attitude-congruent over attitude-incongruent information.

Cognitive accounts of belief bias draw on the notion of Bayesian belief updating in that high confidence in one's beliefs (i.e., strong Bayesian priors) may enhance the tendency to readily accept belief-congruent information and reject belief-incongruent information (Ditto et al., in press; Gawronski, 2021; Gawronski et al., 2023; Pennycook & Rand, 2021a; Tappin et al., 2020). Conversely, weak confidence in one's beliefs (i.e., weak Bayesian priors) should attenuate differences in the acceptance of belief-congruent and belief-incongruent information. Together, these assumptions suggest that high levels of confidence in one's beliefs should be positively associated with belief bias favoring attitude-congruent over attitude-incongruent information.

The main goal of Experiment 3 was to disentangle the two accounts by testing their unique predictions about associations of belief bias with feelings about oneself (*self-feelings*) and confidence in one's beliefs (*self-confidence*). Specifically, we tested whether positive self-feelings are negatively associated with belief-congruency bias (as predicted by motivational accounts), and whether self-confidence is positively associated with belief-congruency bias (as predicted by cognitive accounts). We did not expect self-feelings and self-confidence to be associated with truth sensitivity. Because Experiment 3 is concerned with belief bias favoring attitude-congruent over attitude-incongruent information, we did not recruit participants with neutral attitudes toward COVID-19 vaccines in this study.

**Methods**

*Participants and Design*

Data for Experiment 3 were collected in September 2022. We used the crowdsourcing platform Prolific and its prescreening data to separately recruit 200 participants who had reported feeling positively about COVID-19 vaccines and 200 participants who had reported feeling negatively about COVID-19 vaccines. The recruitment was based on responses to Prolific's prescreening question *Please describe your attitudes towards the COVID-19 (Coronavirus) vaccines* with the four response options (1) *For (I feel positively about the vaccines)*, (2) *Against (I feel negatively about the vaccines)*, (3) *Neutral (I don't have strong opinions either way)*, (4) *Prefer not to say*. In addition to using the recruitment filters of Experiment 1, we restricted participation to Prolific workers who had not participated in Experiments 1 and 2. The experiment took approximately 10-15 minutes. Participants were compensated US-$3 for their time. The experiment utilized a 2 (Participant Attitude: favorable vs. unfavorable) × 2 (Statement Accuracy: true vs. false) × 2 (Statement Slant: pro-COVID-19-vaccine vs. anti-COVID-19-vaccine) mixed design with the first factor varying between subjects and the last two factors varying within subjects.

When determining the desired sample size, we anticipated that approximately 10% of participants would be excluded from analyses based on preregistered exclusion criteria (see below). For the preregistered 2 (Participant Attitude, between-subjects) × 2 (Statement Slant, within-subjects) mixed ANOVAs, a sample of $N = 360$ (90% of the sample) provides a power of 80% to detect small effects of $f = 0.119$ for the main effect of Participant Attitude and $f = 0.088$ for the main effect of Statement Slant and the Participant Attitude × Statement Slant interaction. For the preregistered multiple-regression analyses with two predictors, a sample of $N = 360$ provides a power of 80% to detect a small effect of $f^2 = 0.022$ of one predictor.

As preregistered, we ended data collection after 400 participants had been approved credit on Prolific. The number of cases with complete submissions was 400. We used two preregistered criteria to exclude participants with complete submissions from the analyses. First, we excluded 78 participants who failed our attention check.[15] Second, we excluded 32 participants who reported inconsistent attitudes toward COVID-19 vaccines in Prolific's prescreening survey and the measure of attitudes toward COVID-19 vaccines in our experiments.[16] The remaining sample of 290 participants included 164 participants with favorable attitudes and 126 participants with unfavorable attitudes. Of the 290 participants in the final sample, 147 identified as female, 135 as male, and 8 chose the response option *other*. The age range was 18 to 77 years ($M_{age}$ = 40.38, $SD_{age}$ = 13.97). Seven of the retained participants indicated being American Indian or Alaska Native, 18 Asian, 19 Black, 19 Hispanic, Latino, or Spanish origin, 3 Middle Eastern or North African, 234 White, and 6 chose the response option *other*. Of the retained participants, 9 reported having less than a high school diploma or equivalent, 133 a high school diploma or equivalent, 111 an associate or bachelor's degree, 34 a master's degree, and 3 a doctoral degree. Regarding country of residence, 158 participants reported currently residing in the UK and 132 in the US. For the preregistered 2 (Participant Attitude, between-subjects) × 2 (Statement Slant, within-subjects) mixed ANOVAs, the final sample of $N$ = 290 provides 80% power to detect small effects of $f$ = 0.133 for the main effect of Participant Attitude and $f$ = 0.098 for the main effect of Statement Slant and the Participant Attitude × Statement Slant interaction. For the preregistered multiple-regressions with two predictors, the final sample of $N$ = 290 provides 80% power to detect a small effect of $f^2$ = 0.027 of one predictor.

### Materials, Procedure, and Measures

The materials, procedure, and measures were identical to Experiment 1, the only two differences being that (*i*) the response option *two booster doses* in the question about COVID-19 vaccination status was changed to *two booster doses or more* and (*ii*) we measured participants' self-feelings and self-confidence prior to the statement-judgment task. To measure self-feelings, participants rated how positive, negative, good, and bad they felt about themselves on four scales ranging from *Not at all positive* (1) to *Extremely positive* (5), from *Not at all negative* (1) to *Extremely negative* (5), from *Not at all good* (1) to *Extremely good* (5), and from *Not at all bad* (1) to *Extremely bad* (5), respectively. To measure self-confidence, participants rated how confident, unconfident, certain, and uncertain they felt about their personal views on four scales ranging from *Not at all confident* (1) to *Extremely confident* (5), from *Not at all unconfident* (1) to *Extremely unconfident* (5), from *Not at all certain* (1) to *Extremely certain* (5), and from *Not at all uncertain* (1) to *Extremely uncertain* (5), respectively. The order of the two measures was counterbalanced between participants, and the order of the items for each construct was randomized for each participant. Each item was presented on a separate screen.

### Results

As preregistered, data aggregation followed the procedures outlined in Experiment 1, the only preregistered difference being that we additionally calculated an index of belief-congruency bias as the difference between acceptance thresholds for attitude-congruent versus attitude-incongruent statements. Higher scores on this index reflect a lower threshold for accepting attitude-congruent compared to attitude-incongruent statements.

Non-preregistered descriptive analyses revealed that participants showed a high ability to differentiate between true and false information about COVID-19 vaccines, in that the average $d'$ score at the sample level was positive and significantly different from zero ($M$ = 1.49, $SD$ = 0.73), $t(289)$ = 34.66, $p < .001$, $d$ = 2.04. Further analyses revealed that, on average, participants were more likely to reject than to accept information about COVID-19 vaccines, which is reflected in a positive $c$ score that significantly differed from zero ($M$ = 0.19, $SD$ = 0.23), $t(289)$ = 14.14, $p < .001$, $d$ = 0.83. At the sample level, participants showed an anti-COVID-19-vaccine belief bias, in that $c$ scores were significantly lower for anti-COVID-19-vaccine than pro-COVID-19-vaccine information ($M$s = -0.08 vs. 0.48, respectively), $F(1, 288)$ = 132.48, $p < .001$, $\eta^2_p$ = .32. Replicating the pattern obtained in Experiments 1 and 2, truth sensitivity and anti-COVID-19-vaccine belief bias were negatively correlated, in that greater truth sensitivity was associated with a lower anti-COVID-19-vaccine belief bias ($r$ = -.59, $p < .001$).[17]

### Truth Sensitivity

Following our preregistered analysis plan, $d'$ scores were submitted to a 2 (Participant Attitude) × 2 (Statement Slant) mixed ANOVA with the first variable as a between-subjects factor and the second variable as a within-subjects factor. A significant main effect of Participant Attitude indicated that participants with favorable attitudes toward COVID-19 vaccines showed a higher ability to discern true from false information about COVID-19 vaccines compared to participants with unfavorable attitudes, $F(1, 288) = 153.54$, $p < .001$, $\eta^2_p = .35$ (see Figure 1, bottom-left panel), replicating the results of Experiments 1 and 2. Non-preregistered exploratory analyses revealed that the main effect of Participant Attitude remained statistically significant after controlling for participants' country of residence, gender, racial/ethnic identity, political orientation, age, and education, $F(1, 275) = 90.38$, $p < .001$.

---

[15] To explore potential effects of selective attrition (Zhou & Fishbach, 2016), we also ran all preregistered analyses without excluding participants who failed the attention check. All findings focal to our main research questions replicated in these analyses. Two non-focal differences occurred, in that (*i*) participants showed higher truth sensitivity for pro- compared to anti-COVID-19-vaccine information and (*ii*) the negative association between self-confidence and truth sensitivity was non-significant ($\beta$ = -.100, $p$ = .075).

[16] Exploratory analyses examining truth sensitivity and belief bias among the small number of participants who reported inconsistent attitudes are reported in the Supplemental Materials.

[17] Non-preregistered exploratory analyses on demographic differences are reported in the Supplemental Materials.

A significant two-way interaction between Participant Attitude and Statement Slant further indicated that participants with different attitudes toward COVID-19-vaccines differed in terms of their truth sensitivity for pro- and anti-COVID-19-vaccine information, $F(1, 288) = 7.30$, $p = .007$, $\eta^2_p = .02$ (see Figure 1, bottom-left panel). Replicating the results of Experiments 1 and 2, participants with favorable attitudes showed greater truth sensitivity for pro- compared to anti-COVID-19-vaccine information, $t(163) = 3.48$, $p = .001$, $d_z = 0.27$. Different from the results of Experiment 1, but consistent with the results of Experiment 2, truth sensitivity for the two types of information did not significantly differ among participants with unfavorable attitudes, $t(125) = -0.52$, $p = .605$, $d_z = 0.05$.

Nevertheless, the pattern of differences between the two attitude groups was consistent across the two types of statements, replicating the results of Experiments 1 and 2. Specifically, truth sensitivity was greater among participants with favorable attitudes compared to participants with unfavorable attitudes for both pro-COVID-19-vaccine information, $t(288) = 12.34$, $p < .001$, $d = 1.46$, and anti-COVID-19-vaccine information, $t(288) = 9.13$, $p < .001$, $d = 1.08$.

*Acceptance Threshold*

Following our preregistered analysis plan, $c$ scores were submitted to a 2 (Participant Attitude) × 2 (Statement Slant) mixed ANOVA with the first variable as a between-subjects factor and the second variable as a within-subjects factor. A significant main effect of Participant Attitude revealed that participants with unfavorable attitudes had a significantly higher acceptance threshold than participants with favorable attitudes, $F(1, 288) = 11.50$, $p = .001$, $\eta^2_p = .04$ (see Figure 1, bottom-right panel). This main effect was qualified by a significant two-way interaction between Participant Attitude and Statement Slant, indicating that participants with different attitudes toward COVID-19 vaccines differed in their acceptance thresholds for pro- versus anti-COVID-19-vaccine information, $F(1, 288) = 473.60$, $p < .001$, $\eta^2_p = .62$ (see Figure 1, bottom-right panel). Replicating the results of Experiments 1 and 2, participants with favorable attitudes had a significantly lower acceptance threshold for pro-COVID-19-vaccine information than anti-COVID-19-vaccine information, and thus showed a pro-COVID-19-vaccine belief bias, $t(163) = -10.56$, $p < .001$, $d_z = 0.82$. Conversely, participants with unfavorable attitudes had a significantly lower acceptance threshold for anti-COVID-19-vaccine information than pro-COVID-19-vaccine information, and thus showed an anti-COVID-19-vaccine belief bias, $t(125) = 17.50$, $p < .001$, $d_z = 1.56$.[18] Non-preregistered exploratory analyses revealed that the interaction between Participant Attitude and Statement Slant remained statistically significant after controlling for country of residence, gender, racial/ethnic identity, political orientation, age, and education, $F(1, 288) = 473.60$, $p < .001$.

*Acceptance of False Information*

Preregistered confirmatory analyses regressing acceptance of false information (i.e., false-alarm rate) onto overall truth-sensitivity and anti-COVID-19-vaccine belief bias as simultaneous predictors followed the procedures outlined in Experiment 1. The results of the multiple-regression analyses are presented in Table 2. Replicating the results of Experiments 1 and 2, truth sensitivity showed a reliable negative association with acceptance of false information about COVID-19 vaccines regardless of whether the false information had a pro- or anti-COVID-19-vaccine slant. Also replicating the results of Experiments 1 and 2, anti-COVID-19-vaccine belief bias showed a reliable positive association with acceptance of anti-COVID-19-vaccine misinformation and a reliable negative association with acceptance of pro-COVID-19-vaccine misinformation. As with Experiments 1 and 2, standardized regression coefficients for the obtained associations with belief bias were approximately twice as large as those obtained for truth sensitivity. All effects replicated regardless of whether the outcome variable was based on odd-numbered items and the predictors based on even-numbered items, or vice versa.[19]

*Self-Feelings and Self-Confidence*

Indices of self-feelings and self-confidence were calculated by reverse coding the two negatively-framed items for each construct and then computing the mean across the four items of each construct (self-feelings: $\alpha = .94$; self-confidence: $\alpha = .81$). Higher values on these indices reflect more positive self-feelings and greater confidence in one's beliefs, respectively. Self-feelings and self-confidence were positively correlated, in that more positive self-feelings were associated with higher levels of confidence in one's beliefs ($r = .33$, $p < .001$).

---

[18] Follow-up tests comparing acceptance thresholds of the two Participant Attitude groups for pro-vaccine versus anti-vaccine information, respectively, are reported in the Supplemental Materials.

[19] Because truth sensitivity and anti-COVID-19-vaccine belief bias were negatively correlated, we computed variance inflation factors (VIFs) to rule out potential problems with multicollinearity. The VIFs suggest that multicollinearity is not problematic in the current data; predictors even-numbered items: VIF = 1.49, predictors odd-numbered items: VIF = 1.58.

Table 3 depicts the results of the preregistered confirmatory analyses using self-feelings and self-confidence as simultaneous predictors of truth sensitivity and belief-congruency bias, respectively. Consistent with the prediction derived from cognitive accounts, self-confidence showed a significant positive association with belief-congruency bias, indicating that higher levels of confidence were associated with a stronger tendency to accept attitude-congruent information and reject attitude-incongruent information. Yet, contrary to the prediction derived from motivational accounts, self-feelings showed no significant association with belief-congruency bias. Non-preregistered exploratory analyses revealed that belief-congruency bias was significantly associated with self-confidence (but not self-feelings) after controlling for participants' country of residence, gender, racial/ethnic identity, political orientation, age, and education.

As hypothesized, self-feelings showed no significant association with truth sensitivity. However, contrary to our hypothesis, self-confidence showed a significant negative association with truth sensitivity, indicating that higher levels of confidence were associated with a lower ability to discern true from false information about COVID-19 vaccines. Non-preregistered exploratory analyses revealed that truth sensitivity was significantly associated with self-confidence (but not self-feelings) after controlling for participants' country of residence, gender, racial/ethnic identity, political orientation, age, and education.

**Discussion**

Experiment 3 obtained three sets of important findings. First, preregistered confirmatory analyses replicated the differences between vaccine-attitude groups obtained in Experiments 1 and 2, corroborating the reliability of these differences. Second, preregistered confirmatory analyses replicated the predictive relations obtained in Experiment 1, in that acceptance of false information about COVID-19 vaccines was jointly predicted by both truth sensitivity and belief bias, with belief bias being the stronger predictor. Third, preregistered confirmatory analyses revealed a significant positive association between self-confidence and belief-congruency bias, supporting predictions derived from cognitive accounts of belief bias. Yet, counter to predictions derived from motivational accounts of belief bias, positive self-feelings did not show a significant negative association with belief-congruency bias. Unexpectedly, self-confidence also showed a significant negative association with truth sensitivity. However, because this association did not reach statistical significance in exploratory robustness analyses of our data (see Footnote 15), we refrain from interpreting this unexpected finding.

**General Discussion**

A large body of research has provided valuable insights into psychological factors underlying susceptibility to misinformation (for reviews, see Ecker et al., 2022; van der Linden, 2022). However, a major limitation of this work is the use of methodological approaches that focus exclusively on the ability to distinguish between true and false information, which has led to confusion about the contribution of belief bias to misinformation susceptibility. To address this limitation, the current work used SDT to investigate the role of truth sensitivity and belief bias in judgments of (mis)information about COVID-19 vaccines.

Overall, participants in the current studies performed extremely well in discerning true from false information (Cohen's $d$s for positive $d'$ scores = ~2.0) and they were very cautious in accepting information as true (Cohen's $d$s for positive $c$ scores = ~0.85). Nevertheless, participants also accepted a considerable amount of false information, with acceptance of misinformation being jointly predicted by truth insensitivity and belief bias. While truth insensitivity increased the risk of accepting false information regardless of the statement's slant, belief bias functioned as either a risk or protective factor, depending on the direction of the bias and the slant of misinformation. Whereas anti-COVID-19-vaccine belief bias increased acceptance of false anti-COVID-19-vaccine information and reduced acceptance of false pro-COVID-19-vaccine information, pro-COVID-19-vaccine belief bias increased acceptance of false pro-COVID-19-vaccine information and reduced acceptance of false anti-COVID-19-vaccine information. Although acceptance of false information was jointly predicted by both truth insensitivity and belief bias, the obtained associations with belief bias were larger, with effect sizes that were about twice the size of the obtained associations with truth insensitivity.

Attitudes toward COVID-19 vaccines emerged as a major correlate of both truth insensitivity and belief bias. Among the different groups of participants, those with unfavorable attitudes toward COVID-19 vaccines showed the lowest truth sensitivity and a large anti-COVID-19-vaccine belief bias, which renders this group particularly vulnerable to anti-COVID-19-vaccine misinformation. Although participants with neutral attitudes performed better in distinguishing between true and false information than participants with unfavorable attitudes, they still performed worse than participants with favorable attitudes. Participants with neutral attitudes also showed a considerable anti-COVID-19-vaccine belief bias. Participants with favorable attitudes toward COVID-19 vaccines showed the highest levels of truth sensitivity for both pro-COVID-19-vaccine and anti-COVID-19-vaccine information. However, this group of participants also showed a strong belief bias, in that they were more likely to accept information with a pro-COVID-19-vaccine slant than information with an anti-COVID-19-vaccine slant. Although belief-congruency bias was smaller among participants with favorable attitudes than among participants with unfavorable attitudes, a pro-COVID-19-vaccine belief bias can increase susceptibility to misinformation by leading people to accept false pro-COVID-19-vaccine information as true.[20]

Consistent with the results of prior research (e.g., Bago et al., 2020; Gawronski et al., 2023; Pennycook & Rand, 2019; Sultan et al., 2022), we found that truth sensitivity was weaker under conditions that interfere with cognitive elaboration (i.e., time pressure). However, unexpectedly, this effect was limited to anti-COVID-19-vaccine information and did not generalize to pro-COVID-19-vaccine information. These findings suggest that thoughtful processing may help to distinguish real concerns from false claims in messages about negative aspects of COVID-19 vaccines. However, thoughtful processing may be less effective in increasing the ability to distinguish between true and false claims in messages about positive aspects of COVID-19 vaccines. Importantly, although greater cognitive elaboration increased truth sensitivity to some extent, it was ineffective in reducing belief bias. Because belief bias explained much larger portions of variance in acceptance of false information than truth sensitivity, accounts that attribute misinformation susceptibility to insufficient cognitive elaboration (e.g., Pennycook, 2023; Pennycook & Rand, 2021a) are missing an important factor underlying acceptance of false information.

Counter to predictions derived from motivational accounts of belief-congruency bias (see Ditto et al., in press), we did not find evidence for a negative association between positive self-feelings and belief-congruency bias. According to motivational accounts, people who feel more positive about themselves should be less prone to showing a belief-congruency bias because they do not need to regulate their self-feelings by seeking self-validation or avoiding self-threat (Gawronski et al., 2023; Liberman & Chaiken, 1992; Sherman et al., 2000). Conversely, people who feel less positive about themselves should be more prone to showing a belief-congruency bias because doing so helps to elevate positive feelings about the self. In the current work, we did not find any support for these assumptions.

Consistent with predictions derived from cognitive accounts of belief-congruency bias (Pennycook & Rand, 2021a; Tappin et al., 2020), high confidence in one's views was associated with a stronger belief-congruency bias. According to cognitive accounts invoking principles of Bayesian belief updating, high confidence in one's beliefs functions in a manner akin to strong Bayesian priors, leading people to accept belief-congruent information and reject belief-incongruent information. Conversely, weak confidence in one's beliefs functions in a manner akin to weak Bayesian priors, which should attenuate differences in the acceptance of belief-congruent and belief-incongruent information. These assumptions suggest that high levels of self-confidence should be positively associated with belief bias favoring attitude-congruent over attitude-incongruent information, as found in the current research.

**Theoretical Implications**

The current findings stand in stark contrast to prominent claims that belief-congruency bias in responses to misinformation may be negligible or even non-existent (Pennycook & Rand, 2019, 2021a, 2021b). What is more, our results question the relative importance of people's ability to discern true from false information for the propensity to accept misinformation, challenging the claim that truth sensitivity plays a crucial role in misinformation susceptibility (e.g., Pennycook & Rand, 2021a, 2021b).

Examining the psychological processes underlying belief-congruency bias, we did not find support for a prevailing narrative that attributes belief-congruency bias to processes of motivated reasoning arising from a desire to protect and support one's personal beliefs (see Ditto et al., in press). Instead, we found evidence for recent speculations that belief-congruency bias may be the product of cognitive processes following the principles of Bayesian belief updating (Pennycook & Rand, 2021a; Tappin et al., 2020), with people's confidence in their pre-existing views corresponding to the strength of Bayesian priors. However, while this account explains the belief-congruency biases of participants with unfavorable and favorable attitudes as well as the obtained association between belief-congruency bias and self-confidence, it does not explain the anti-COVID-19-vaccine belief bias among participants with neutral attitudes. Because neutral attitudes toward COVID-19 vaccines are neither congruent nor incongruent with either positive or negative information about COVID-19 vaccines, it remains unclear why participants with neutral attitudes toward COVID-19 vaccines showed a tendency to accept negative and reject positive information about the vaccines in all three studies. One possibility is that participants with neutral attitudes are more strongly influenced by negative information, as suggested by recent work on political attitudes (Siev et al., 2024). Further research may examine why individuals who hold neutral attitudes exhibit a tendency to accept information with a particular slant and reject information with the opposite slant. Regarding alternative motivational underpinnings of belief-congruency bias, future research may also investigate other motivational drivers such as need to belong (Rathje et al., 2023) or need for chaos (Arceneaux et al., 2021).

---

[20] An interesting question is why participants with favorable attitudes toward COVID-19 vaccines showed higher truth sensitivity compared to participants with neutral or unfavorable attitudes. One potential reason is that higher truth sensitivity leads people to develop more favorable attitudes toward COVID-19 vaccines. Another (not mutually exclusive) reason is that a favorable attitude toward COVID-19 vaccines leads people to search for information from trustworthy sources, which in turn increases truth sensitivity.

The current research adds to a growing line of work showing the value of SDT as a framework for understanding susceptibility to misinformation (see Batailler et al., 2022). Using SDT to investigate the role of truth sensitivity and partisan bias in responses to political (mis)information, Gawronski et al. (2023) found strong partisan-bias effects in both judgments of truth and decisions to share information. Moreover, although participants showed much higher thresholds for sharing information than judging information as true, the higher thresholds for sharing decisions did not lead to greater accuracy, in that truth sensitivity was lower (not higher) for sharing decisions than judgments of truth. Expanding on prior work on cognitive elaboration (Bago et al., 2020; Pennycook & Rand, 2019), Sultan et al. (2022) found that time pressure reduced truth sensitivity in judgments of political (mis)information without affecting acceptance thresholds for ideology-congruent and ideology-incongruent information (see also Gawronski et al., 2023). Using SDT to reanalyze data from studies testing the effectiveness of gamified interventions to reduce misinformation susceptibility (e.g., Basol et al., 2021; Roozenbeek & van der Linden, 2019), Modirrousta-Galian and Higham (2023) found that the tested interventions were largely ineffective in increasing participants' ability to distinguish between true and false information. Instead, the interventions merely increased participants' thresholds for accepting information as true.

The current research adds to this body of work in at least four ways. First, the current research goes beyond prior applications of SDT to political (mis)information by investigating the role of truth sensitivity and belief bias in judgments of health-related (mis)information.[21] Second, the current findings corroborate prior conclusions that, although greater cognitive elaboration is effective in increasing truth sensitivity, it is ineffective in reducing belief bias. Third, the current findings pose a challenge to the dominant claim that belief bias is irrelevant for understanding susceptibility to misinformation. Fourth, the current findings pose a challenge to accounts that attribute belief bias to processes of motivational reasoning, and instead suggest that belief bias might be the product of cognitive processes that conform to the principles of Bayesian belief updating. Together, these contributions provide further support for the value of SDT as a framework for understanding susceptibility to misinformation. Based on this conclusion, we suggest that future research on misinformation susceptibility should adopt SDT as a general framework instead of relying on approaches that focus exclusively on truth discernment.

**Practical Implications**

In addition to its theoretical contributions, the current work also has important practical implications for attempts to combat misinformation about vaccines. Specifically, the current findings highlight why it can be difficult to convince vaccine skeptics by providing them with positive information about vaccine effectiveness and safety. In the current studies, participants with neutral and unfavorable attitudes toward COVID-19 vaccines both showed a strong anti-COVID-19-vaccine belief bias, in that they more readily accepted anti-COVID-19-vaccine than pro-COVID-19-vaccine information regardless of whether the information was true or false. In other words, both groups showed a tendency to accept false negative information about COVID-19 vaccines as true and to dismiss true positive information about COVID-19 vaccines as false. Such a tendency can bolster unfavorable attitudes toward vaccines and create psychological immunity against efforts to improve vaccine attitudes via positive information about their effectiveness and safety.

Regarding interventions that aim to reduce susceptibility to misinformation, the current findings suggest that nudging people to slow down when scrolling through news and social media might be a potential strategy to increase people's ability to distinguish between true and false information (Kahneman, 2011; Pennycook & Rand, 2021a). However, the current findings also suggest that the impact of such interventions may be relatively limited because (*i*) the effect of processing time on truth sensitivity was rather small in terms of current conventions (Cohen, 1988) and (*ii*) the association between truth sensitivity and acceptance of false information was much smaller compared to the relatively large association between belief bias and acceptance of false information. A more promising approach might be to target people's confidence in their beliefs via interventions to increase intellectual humility (Porter et al., 2022). In the current work, high levels of confidence in one's beliefs were associated with greater belief bias, which suggests that greater intellectual humility might reduce susceptibility to misinformation by tackling this risk factor. Although our correlational findings regarding self-confidence do not permit causal inferences about the impact of interventions that aim to increase intellectual humility, preliminary evidence for the effectiveness of such interventions comes from studies showing that experimental manipulations to increase intellectual humility can reduce susceptibility to political misinformation (Koetke et al., 2023).

---

[21] Although debates about COVID-19 vaccines are highly politicized in the United States (Van Bavel et al., in press), the current findings are independent of political partisanship in that all focal effects replicated after controlling for participants' political orientation (and various other demographic variables).

The finding that participants were more likely to reject than to accept COVID-19-vaccine information (i.e., they showed high threshold scores overall) suggests that, overall, people tend to be more skeptical than gullible. However, while a general tendency to reject information as false acts as a protective factor against accepting misinformation, it leads people to reject true information as false (see also Pfänder & Altay, 2023). Yet, inaccurate beliefs can be rooted in either acceptance of false information or rejection of true information, and the two sources of inaccurate beliefs likely require different types of interventions.

Although the current research focused specifically on misinformation about COVID-19 vaccines, the obtained results also have important implications for other societal challenges. For example, our findings suggest two potential reasons why providing the public with scientific evidence about climate change might have limited impact in fighting climate misinformation. First, similar to people who hold unfavorable attitudes toward COVID-19 vaccines, climate-change skeptics may be difficult to persuade with informational campaigns, because climate-change skeptics may reject belief-incongruent information when judging information as true or false. Second, as with susceptibility to misinformation about COVID-19 vaccines, inability to discern true from false information about climate change may not be the core problem. Instead, a belief bias bolstering pre-existing views may play a much stronger role, in that it leads climate-change skeptics to accept false claims that climate change does not exist and to reject true information about the significance of climate change. Thus, interventions that tackle belief bias will presumably be more effective in convincing climate-change skeptics than interventions targeting truth insensitivity. While these assumptions remain speculative in the absence of direct empirical evidence, they are consistent with earlier conclusions suggesting that belief bias regarding climate change may leave informational efforts fruitless and that different types of interventions may be needed (Druckman & McGrath, 2019). Because belief bias has been found to play a similarly important role in other areas (Gawronski et al., 2023), our conclusions about its significance may apply to a wide range of societally challenging topics.

**Limitations and Constraints on Generality**

In line with our preregistered exclusion criteria, we excluded participants who (*i*) failed to pass a reading-intensive check or (*ii*) provided inconsistent reports about their COVID-19-vaccine attitudes in Prolific's prescreening and the measure in our studies. The first criterion aimed to ensure high data quality; the second criterion was necessary for theoretical reasons to ensure the validity of our manipulations and measures. Yet, in conjunction, the two criteria led to rather high attrition rates. We addressed concerns about high attrition in three ways. First, we exploratorily reran all preregistered analyses including participants regardless of their attention-check response (and the number of missing values due to the 7-second time limit in Experiment 2). None of the findings focal to our main research questions changed. Three non-focal differences occurred, which we describe in Footnotes 9 and 15. Second, we report exploratory analyses examining truth sensitivity and belief bias among participants who reported inconsistent attitudes toward COVID-19 vaccines in Prolific's prescreening survey and the demographic survey in our experiments in the Supplemental Materials. Third, we report sensitivity power analyses for the analyzed sample sizes. All analyzed sample sizes were suitable to detect small effects with a power of 80%.

Although the samples of participants in the current studies were quite diverse in terms of racial/ethnic identity, gender, age, and educational level, a notable limitation is that we exclusively recruited participants who resided in the United States or the United Kingdom. Interestingly, despite large differences between the two countries' response to the COVID-19 pandemic, there were no reliable differences between the two countries regarding truth sensitivity or acceptance thresholds, and country was not a reliable moderator of the association between COVID-19-vaccine attitudes and truth sensitivity or belief bias across the three experiments. While these findings increase our confidence in the generalizability of the results across different contexts, the sub-samples were rather small, implying that they may have been underpowered for the detection of country-level effects. Furthermore, because COVID-19-vaccine information is at least partly country-specific (e.g., authorized vaccines), the reliance on samples from two specific countries could potentially undermine the generalizability of the obtained results to other regions. While this constraint on generalizability is somewhat mitigated by the fact that misinformation poses similar problems and calls for a similar fightback in countries all over the world (Porter & Wood, 2021; Roozenbeek et al., 2020), future research with participants from other countries would be helpful to corroborate our conclusions about truth sensitivity and belief bias in responses to (mis)information about vaccines.

Another characteristic of our samples is that we recruited all participants on Prolific. Although it would be valuable to replicate the current results with a different type of sample, Prolific has been shown to provide high-quality data for behavioral research when compared to other platforms and panels and outperforms more expensive online panels such as Qualtrics Panels and Dynata on key data quality measures (Douglas et al., 2023; Peer et al., 2022).

## Conclusion

The current work investigated why people accept misinformation about COVID-19 vaccines. To this end, we used SDT to quantify two factors that can make people susceptible to accepting false information as true: (*i*) inability to discern true from false information (*truth insensitivity*) and (*ii*) a tendency to accept information with a particular slant regardless of whether it is true or false (*belief bias*). The current findings suggest that belief biases associated with prior attitudes are a major driver of why individuals accept misinformation, while inability to differentiate between true and false information plays a comparatively minor role. Moreover, belief biases associated with vaccine attitudes can make it difficult to convince skeptics via the provision of positive information about the effectiveness and safety of vaccines. Although interventions to promote slow processing might help to increase truth sensitivity, the overall impact of such interventions is likely limited because (*i*) the effect of processing time on truth sensitivity is rather small and (*ii*) truth sensitivity played a much weaker role in acceptance of false information than belief bias. Because having high confidence in one's beliefs was associated with a stronger belief bias, a more effective way to combat misinformation might be with interventions to increase intellectual humility. Although the effectiveness of such interventions remains to be tested, the current research provides valuable insights for this endeavor by revealing why people fall for misinformation in the domain of COVID-19 vaccines.

## Author Note

The authors declare no conflict of interest. We thank Doonya Tabibi for her help with the preparation of the study materials. This research was supported by National Science Foundation Grant BCS-2040684, Swiss National Science Foundation Grant P500PS_214298, and the German Academic Exchange Service. Any opinions, findings, and conclusions or recommendations expressed in this material are those of the authors and do not necessarily reflect the views of the funding agencies.

Correspondence concerning this article should be addressed to Lea S. Nahon, Department of Psychology, University of Texas at Austin, 108 E. Dean Keeton A8000, Austin, TX 78712. Email: lea.nahon@utexas.edu

## CRediT Authorship Contribution Statement

Lea S. Nahon: Conceptualization, Methodology, Software, Validation, Formal analysis, Investigation, Data curation, Writing - original draft, Visualization, Project administration. Nyx L. Ng: Conceptualization, Methodology, Writing - review & editing. Bertram Gawronski: Conceptualization, Methodology, Resources, Writing - review & editing, Supervision, Project administration, Funding acquisition.

## Appendix: Material Selection

To generate a large pool of statements that are either true or false and have either a pro-COVID-19-vaccine or an anti-COVID-19-vaccine slant, we searched through online content (e.g., headlines, articles, posts, fact-checks, webpages, etc.) from various types of sources: news sources (AP News, CNN, Fox News, Fox 8, The New York Times, The Washington Post, Reuters, CNBC, NBC News, CBS News, The Guardian, NPR, Star Tribune, The Herald-Times, Detroit Free Press, The Seattle Times, Washington Examiner, Forbes, Insider, Fortune, The Atlantic, USA Today, GMA, Deseret News, National Geographic, Fierce Pharma, Scrubbing In, BBC, CBC, Tagesschau, NDR, Newslodge); authoritative health sources (CDC, FDA, WHO); fact-checking websites (FactCheck.org, PolitiFact, Snopes, Lead Stories, Health Feedback, Poynter); university, hospital, pharmaceutical company, or professional association websites (Yale News, The Brink, UTHealth, Washington University School of Medicine, AMA, Progress West Hospital, Massachusetts General Hospital, Takeda); research articles; social media (Twitter, Reddit, Instagram, Quora); blog or forum posts. The full database of all identified statements as well as their corresponding webpages are available at https://osf.io/utk69/. The database contains information concerning the veracity of the statements, the source URL, the source type, the publication date, the date the statement was entered into the database, context information about the statement, why the statement is considered true or false, a fact-check link or information, additional notes, and the initials of the person who entered the statement into the database.

In a first screening, we excluded (i.e., marked red) all statements that were about COVID-19, but not about COVID-19 vaccines. We then further screened the statements to select (i.e., mark green) suitable statements. We either excluded (i.e., marked blue) or adapted statements that (*i*) were not unambiguously true or false, (*ii*) were not clearly favorable or unfavorable of COVID-19 vaccines, (*iii*) were overly long or complicated, (*iv*) involved a person or entity claiming something (e.g., *Pfizer says COVID-19 vaccine works in kids ages 5 to 11.*), (*v*) were too extreme or involved conspiracy theories (e.g., *Vaccination will be compulsory which will alter human DNA and will be aimed at universal chipping.*), (*vi*) were outdated, (*vii*) could easily change their truth status in the future, (*viii*) may be too difficult to understand or were not suitable for the study sample at the time of data collection (e.g., statements about Novavax), or (*ix*) were duplicates. We also modified statements to be clearer and easier to read. In a next step, we went through the statements marked green and preselected only those statements for which we had no remaining concerns. We excluded or modified statements that (*i*) were too vague, (*ii*) were ambiguous or unclear, (*iii*) were too specific, (*iv*) were not valenced enough (favorable or unfavorable of COVID-19 vaccines), (*v*) were too extreme, (*vi*) entailed scientific jargon, (*vii*) we were unable to fact-check, (*viii*) were challenged by new evidence, (*ix*) were opinion-like or subjective, (*x*) included AstraZeneca because it was not approved in the United States, (*xi*) were not about the vaccines per se (e.g., *14 California children given wrong amount of COVID vaccine.*), or (*xii*) consisted of two parts. For the false pro-COVID-19-vaccine category, we generated additional statements by modifying some statements of the statement pool. Within the set of preselected statements, we marked all statements that seemed most suitable in terms of being not too general, too specific, too subjective, too complicated, redundant, or not about COVID-19 vaccines per se. If possible, we modified problematic statements.

We then selected suitable statements to generate sets of four matching statements, with one statement for each category (i.e., false pro-COVID-19 vaccine, false anti-COVID-19 vaccine, true pro-COVID-19 vaccine, and true anti-COVID-19 vaccine). We aimed to match the four statements within each set in terms of generality, extremity of valence, and content (if possible), and to that end adapted some of the statements. We also added additional statements. Furthermore, we aimed to avoid redundant statements within each statement category and direct contradictions between statements in different categories. Statements were selected or modified accordingly. We fact-checked all selected statements that did not come from an authoritative health source (e.g., CDC). If we were unable to sufficiently fact-check a statement, we replaced it. The final set of statements comprised 20 statements per statement category (i.e., 80 statements total). We standardized the final statements with respect to capitalization and punctuation, and implemented small changes such that the statements were clearly about COVID-19 vaccines, clearly true or false, less complicated, not time-sensitive, grammatically correct, and suitable for samples from the United States and the United Kingdom. The final set of statements were used in Experiments 1 to 3 and can be found in Table S2 of the Supplemental Materials. A file with detailed information about the final set of statements is available at https://osf.io/utk69/. This file includes the source URL, the source type, whether the final statement is based on a headline or standard text, the publication date, the date we entered the statement into the database, context information about the statement, fact-check information and fact-check links, additional notes, the initials of the person who entered the statement into the database, the original statement (i.e., as it appeared in an article, fact-check, social-media post, webpage, or video), and the final statement used in Experiments 1 to 3. The file also includes the statements grouped in the matched sets of four, with each row corresponding to one matched statement set.

**Table 1.** *Binary "True" vs. "False" Judgments of True vs. False Information.*

|  | Response *True* | Response *False* |
|---|---|---|
| True Information | HIT | MISS |
| False Information | FALSE ALARM | CORRECT REJECTION |

*Note.* Using signal-detection terminology, a judgment of true information as true can be described as a *HIT*; a judgment of false information as false can be described as a *CORRECT REJECTION*; a judgment of true information as false can be described as a *MISS*; and a judgment of false information as true can be described as a *FALSE ALARM*.

**Table 2.** *Results of Multiple-Regression Analyses Using Truth Sensitivity and Anti-COVID-19-Vaccine Belief Bias as Simultaneous Predictors of Acceptance of Anti- and Pro-COVID-19-Vaccine Misinformation, Experiments 1-3.*

|  | *N* | Acceptance of Anti-COVID-19-Vaccine Misinformation | | | | Acceptance of Pro-COVID-19-Vaccine Misinformation | | | |
|---|---|---|---|---|---|---|---|---|---|
|  |  | Truth Sensitivity | | Belief Bias | | Truth Sensitivity | | Belief Bias | |
|  |  | β | *p* | β | *p* | β | *p* | β | *p* |
| Experiment 1 | 323 |  |  |  |  |  |  |  |  |
|   even-odd |  | -.283 | < .001 | .635 | < .001 | -.222 | < .001 | -.717 | < .001 |
|   odd-even |  | -.315 | < .001 | .599 | < .001 | -.366 | < .001 | -.791 | < .001 |
| Experiment 2 | 348 |  |  |  |  |  |  |  |  |
|   even-odd |  | -.279 | < .001 | .638 | < .001 | -.276 | < .001 | -.720 | < .001 |
|   odd-even |  | -.332 | < .001 | .581 | < .001 | -.252 | < .001 | -.670 | < .001 |
| Experiment 3 | 290 |  |  |  |  |  |  |  |  |
|   even-odd |  | -.364 | < .001 | .591 | < .001 | -.366 | < .001 | -.843 | < .001 |
|   odd-even |  | -.286 | < .001 | .657 | < .001 | -.401 | < .001 | -.744 | < .001 |

*Note.* For predictors based on even-numbered items, the outcome variable was based on odd-numbered items (even-odd), and vice versa (odd-even).

**Table 3.** *Results of Multiple-Regression Analyses Using Self-Feelings and Self-Confidence as Simultaneous Predictors of Truth Sensitivity and Belief-Congruency Bias, Respectively, Experiment 3 (N = 290).*

|  | Truth Sensitivity | | Belief-Congruency Bias | |
|---|---|---|---|---|
|  | β | *p* | β | *p* |
| Without Controlling for Demographics |  |  |  |  |
|   Self-Feelings | -.042 | .497 | .035 | .567 |
|   Self-Confidence | -.161 | .009 | .249 | < .001 |
| Controlling for Demographics |  |  |  |  |
|   Self-Feelings | .016 | .787 | .021 | .742 |
|   Self-Confidence | -.174 | .002 | .234 | < .001 |

*Note.* Analyses controlling for demographics included country of residence, gender, racial/ethnic identity, political orientation, age, and education as covariates.

**Figure 1.** *Mean Scores of Truth Sensitivity d' (Left Panels) and Acceptance Threshold c (Right Panels) as a Function of Attitudes toward COVID-19 Vaccines (Favorable vs. Neutral vs. Unfavorable) and Information Slant (Pro-COVID-19-Vaccine vs. Anti-COVID-19-Vaccine) in Experiment 1 (Top Panels; N = 323), Experiment 2 (Middle Panels; N = 348), and Experiment 3 (Bottom Panels; N = 290).*

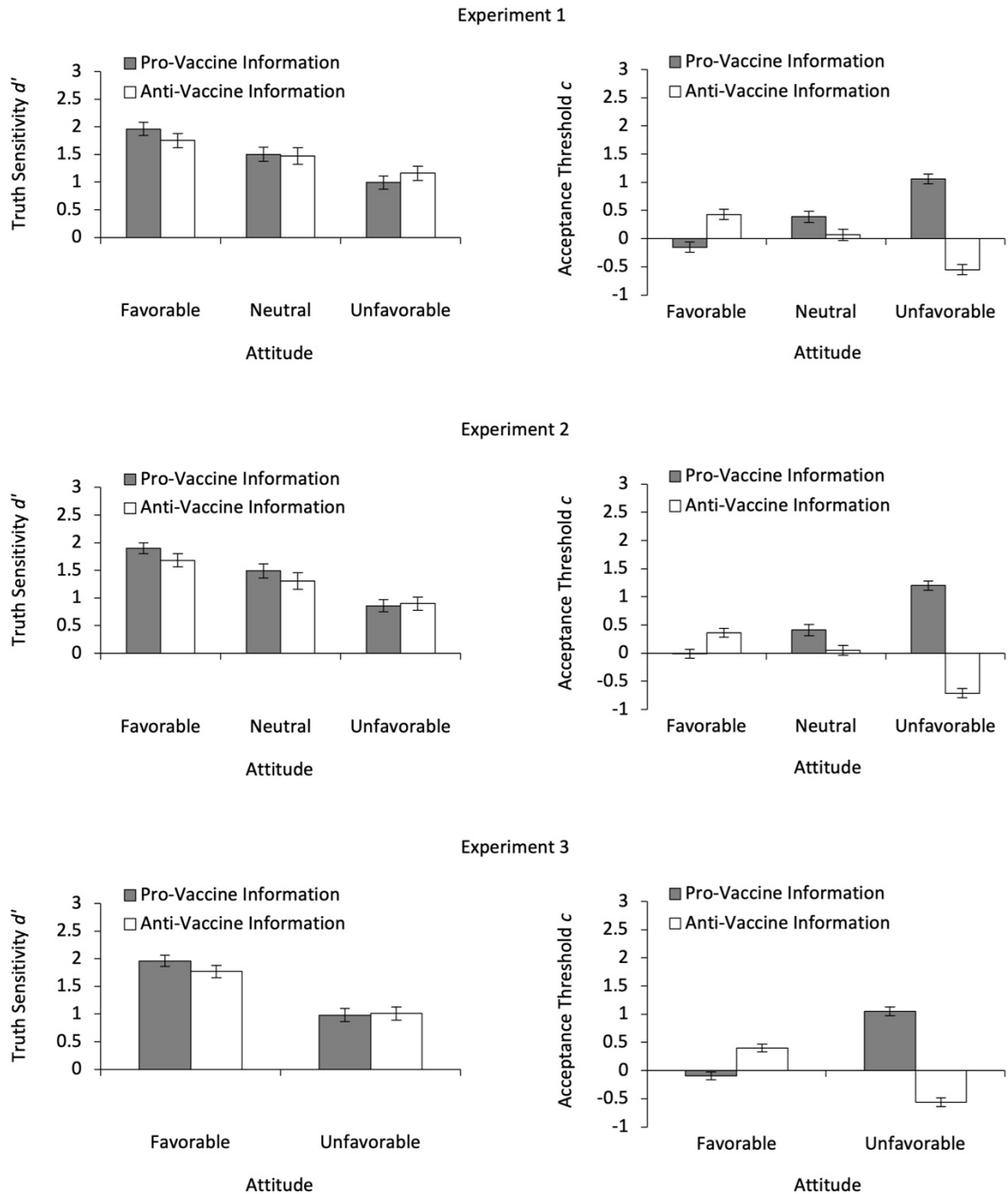

*Note.* Higher truth-sensitivity scores indicate greater ability to accurately distinguish between true and false information about COVID-19 vaccines. Higher acceptance-threshold scores indicate a greater reluctance to accept information as true. Error bars depict 95% confidence intervals.

**Figure 2.** *Mean Truth Sensitivity d' Scores for Pro- and Anti-COVID-19-Vaccine Information as a Function of Cognitive Elaboration (Low vs. High), Experiment 2 (N = 348).*

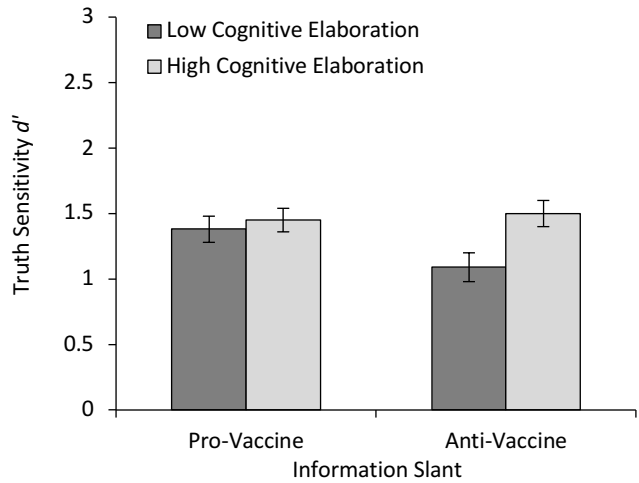

*Note.* Participants in the low-elaboration condition had a 7-second time limit and were asked to provide their answer as quickly as possible. Participants in the high-elaboration condition had unlimited time and were asked to think carefully before providing an answer. Higher scores indicate greater ability to accurately distinguish between true and false information about COVID-19 vaccines. Error bars depict 95% confidence intervals.